\documentclass[12pt]{article}
\usepackage{amsmath}
\usepackage{epsfig}

\begin{document}

\author{D. L. Denny\\
Department of Mathematics\\
James Madison University\\
Harrisonburg VA 22807 \\[6pt]
 and \\[6pt] R. L. Pego\\
Department of Mathematics \&\\
Institute for Physical Science and Technology\\
University of Maryland\\
College Park MD 20742}
\title{Models of low-speed flow for near-critical \\
fluids with gravitational and capillary effects}
\date{February 1998}

\maketitle

\begin{abstract}
We study low-speed flows of a highly compressible, single-phase fluid
in the presence of gravity, for example in a regime appropriate for modeling
recent space-shuttle experiments on fluids near the liquid-vapor
critical point.  In the equations of motion, we include forces due to
capillary stresses that arise from a contribution made by strong
density gradients to the free energy. We derive formally simplified
sets of equations in a low-speed limit analogous to the zero Mach
number limit in combustion theory. 

When viscosity is neglected and gravity is weak, the
simplified system includes: a hyperbolic equation for velocity, a
parabolic equation for temperature, an elliptic equation related to
volume expansion, an integro-differential equation for mean pressure,
and an algebraic equation (the equation of state).  
Solutions are determined by initial values for the mean pressure, 
the temperature field, and the divergence-free part of the velocity field.
To model multidimensional flows with strong gravity, we offer an alternative
to the anelastic approximation, one which 
admits stratified fluids in thermodynamic equilibrium,  
as well as gravity waves but not acoustic waves.

\end{abstract}


\renewcommand{\theequation}{\arabic{section}.\arabic{equation}}

\newcommand{\tgg}{\label}      
\newcommand{\rf}[1]{\ref{#1}}  
\renewcommand{\div}{\nabla\cdot}  
\newcommand{\ENT}{s}
\providecommand{\proof}{\noindent\textbf{Proof:}}
\newcommand{\Z}{{\mathbb Z}}
\newcommand{\R}{{\mathbb R}}
\newcommand{\T}{{\mathbb T}}
\newcommand{\ds}{\displaystyle}
\newcommand{\pnon}{p_1}
\newcommand{\pdim}{\pi}
\newcommand{\pzero}{\bar{p}}
\newcommand{\wbar}{\bar{w}}
\newcommand{\thett}{T}
\newcommand{\theteq}{T_e}
\newcommand{\rhoeq}{\rho_e}
\newcommand{\peq}{p_e}
\newcommand{\newrho}{\check{\rho}}
\newcommand{\Vaisala}{V\"ais\"al\"a\ }
\newcommand{\ddz}[1]{\frac{\partial #1}{\partial z}}
\newcommand{\ddt}[1]{\frac{\partial #1}{\partial t}}
\newcommand{\dd}[2]{\frac{\partial #1}{\partial #2}}
\newcommand{\Tpot}{\theta}

\pagebreak

\section{Introduction}

Near the liquid-vapor critical point, many of the thermophysical
properties of a fluid exhibit a singular behavior. For instance, the
isothermal compressibility and the isobaric thermal expansion
coefficients, as well as the isobaric specific heat, all diverge
strongly at the critical point.  Critical enhancement effects are also
encountered in the behavior of the thermal conductivity and the
viscosity in the vicinity of the critical point, while the thermal
diffusivity approaches zero. These singularities play a major role in
the thermal equilibration of near-critical fluids. 

Understanding the effect of singular fluid properties on dynamics is
not always straightforward.  For example, it has been shown that even
though thermal diffusivity is small, temperature changes in a
near-critical fluid can occur rapidly via a mechanism which causes
adiabatic pressure changes in the bulk of the fluid
\cite{bib14,bib15}.  This adiabatic mechanism creates a strong
coupling between temperature changes occurring at the fluid boundaries
and the temperature response in the interior of the fluid. It works as
follows. A temperature perturbation applied at the boundary of a fluid
causes an expansion in the fluid near the boundary.  Through the
medium of sound waves, this produces an adiabatic pressure change in
the interior of the fluid, and a consequent change in the temperature,
much more rapidly than could be accomplished by thermal diffusion
acting alone. Near the critical point, where the fluid becomes highly
expandable and compressible, the adiabatic mechanism dominates the
early thermal response and creates a `critical speeding-up'
phenomenon. This critical speeding-up has been observed in earth-bound
and low-gravity experiments \cite{bib2,bib4}.

In contrast to the short time-scale of the thermal response,
experimenters have observed a much longer time-scale for the
equilibration of density variations \cite{bib21,IKW2}. Near the
critical point, the divergence of the compressibility and the
influence of gravity can create strong macroscopic density gradients
(upon which microscopic density fluctuations are
superimposed). Although early adiabatic processes act rapidly (within
seconds) to accomplish most of the temperature changes, most of the
relaxation of the density distribution to a new equilibrium state is a
non-adiabatic process driven by the much slower (hours-long) process
of heat diffusion.

Recently, Boukari, Pego, and Gammon \cite{bib3} studied the combined
effects of the adiabatic mechanism and earth's gravity on the
equilibration process in near-critical xenon, using a system of
equations which includes not only the adiabatic effect, but also
one-dimensional fluid motion and heat advection. In numerical
simulations of a temperature step experiment, they found that the
onset of a transient diffusive regime occurs about ten times sooner
than estimated by Onuki, Hao and Ferrell \cite{bib15} in the
zero-gravity case, due to the generation of a small temperature
gradient by the adiabatic pressure quench in the presence of gravity.
Boukari et al.\ also observed that over periods of many hours, no single,
exponentially-decaying mode was ever observed to dominate the
diffusive equilibration process. This conclusion is consistent with
measurements and computations of Zhong and Meyer \cite{bib21} and 
Kogan, Zhong and Meyer \cite{KZM}.

The conclusions drawn in these works were based on results
derived from one-dimensional systems of equations that do not account for
multi-dimensional flows. It is not yet clear how multi-dimensional flows
affect equilibration under gravity.  Zappoli et al.\ \cite{Zetal} have
performed computations of two-dimensional buoyant flow for a van der Waals
fluid fairly near the critical point (1K above critical) where
stratification effects are not yet very large, using an `acoustic
filtering' technique resembling the approach taken in this paper,
and have observed an unusual convection pattern. In the zero-gravity situation,
R.~F. Berg \cite{bib1}  has pointed out certain differences
between one-dimensional and corresponding three-dimensional
results regarding the late diffusive regime.

The purpose of this paper is to systematically derive systems of
multi-dimensional equations which govern the dynamics of a low-speed,
highly compressible, single-phase fluid in the presence of gravity.
We shall include forces due to capillary stresses that arise from a
contribution made to the energy by strong density gradients.  Although
there are no sharp interfaces between phases in equilibrium in the
one-phase regime just above the critical temperature, strong gradients
can be generated as transients \cite{IKW1}.  We anticipate that
nonlocal effects related to the energy of `diffuse interfaces' could
play a big role in generating transient flows in certain circumstances
that are accessible to experimental observation.

Our starting point is the general hydrodynamic equations expressing the
conservation of mass, linear momentum and energy for a compressible fluid
with heat conduction and gravity. We suppose that the 
equation of state is appropriate for conditions near the critical point,
and presume that the fluid is in local equilibrium.  
This implies that the time-scale of interest is longer than a local
relaxation time and that the critical point is not so near that 
the correlation length is macroscopic. These assumptions appear to 
be reasonable for describing the regimes studied in recent experiments.

To account for the influence of steep density gradients on
energy, we adopt a constitutive structure compatible with that described in
the work of J.~E. Dunn and J. Serrin \cite{bib6}. Dunn and Serrin permit
the constitutive quantities (such as the Cauchy stress tensor) to depend
upon the gradient of the density as well as upon density and temperature, so
that capillary effects can now be included in the equations. We account for
such effects in the simplest way, adding a squared gradient term to the
Helmholtz free energy density. The resulting system of equations in
section 2 can be used
to study the influence of hydrodynamics on heat transfer on a time scale
appropriate for resolving sound waves.

Here, however, we are interested in relaxation and flow phenomena that
occur very slowly compared to the time it takes for sound waves to
cross the spatial domain. A systematic procedure for obtaining
simplified equations for compressible flows on long time-scales was
introduced by Rehm and Baum \cite{bib17} and by Majda and Sethian
\cite{bib11} in the context of combustion theory. In section 3 we
apply this procedure to the system at hand. We non-dimensionalize the
equations and estimate the size of the various terms, taking
parameters appropriate to a typical experiment of interest. To obtain
a simplified set of equations, we neglect terms which make the
smallest contribution compared to the other terms.

When the effect of gravity is weak, the result is a simplified model
in which the leading-order pressure is spatially constant, and which
accounts for multi-dimensional fluid motions influenced by the
effects of thermal expansion and contraction, gravitational
compression, thermal diffusion, viscosity and capillarity.
With viscosity included, this simplified model consists of
equations (\rf{15p})--(\rf{15divw}) below, plus the equation of state.

When viscosity is neglected, the simplified system includes 
a hyperbolic equation (for velocity), a parabolic equation (for temperature), 
an elliptic equation (related to volume expansion), an ordinary
integro-differential equation (for mean pressure), and an algebraic equation
(the equation of state). In a forthcoming work \cite{DP2}, we 
prove that the simplified model equations are evolutionary; 
i.e., we show that solutions are determined by suitable initial data. 
The initial data required to determine
the solution consist of the temperature field, the mean pressure, and the
divergence-free part of the velocity field. The density is determined from
the temperature and mean pressure via the equation of state.

Earth's gravity creates a strongly nonlinear density profile in
equilibrium when temperature is close to the critical point, due to 
the high compressibility.
This is not well-modelled by the system (\rf{15p})--(\rf{15divw}).
In section \ref{StrongG} we study how to model multidimensional flows
with strong gravity. If we enforce hydrostatic balance at leading order,
as was done for the one-dimensional case in \cite{bib3}, we find that
generally the flow must remain strictly stratified, and any vertical
fluid motions are due only to horizontally uniform density changes.

In order to admit nontrivial convective flows or gravity waves, one
can assume that entropy is constant at leading order and neglect heat
conduction.  This corresponds to the assumptions made by Ogura and
Phillips \cite{OP} in deriving `anelastic' equations for atmospheric
circulations. Here we obtain anelastic equations valid for a general
equation of state.  The assumption of approximately constant entropy
may not be compatible with thermodynamic equilibrium near the critical
point, but this assumption may be relevant for describing some experiments
that have been performed at near-constant entropy in order to reduce
the effect of density gradients \cite{bib2, bib12}. Because heat conduction
is neglected, though, the anelastic equations do not contain the fast
adiabatic heat-transfer mechanism mentioned above.

We find a better alternative if we return to the weak-gravity scaling
and make a simple modification of the momentum equation.  We retain
the effect of the pressure correction on density in the gravity force
term, in the spirit of the Boussinesq approximation. This modification
does not change the formal validity of the approximation.  But the new
system, consisting of equations (\ref{34t})--(\ref{34.v0}) below,
correctly captures strongly stratified thermodynamic equilibria,
admits multidimensional flow including gravity waves but not acoustic
waves, and includes heat conduction and the adiabatic mechanism.

\section{Basic hydrodynamic equations}\label{Basic}

To start, we consider the general hydrodynamic equations expressing the
conservation of mass, linear momentum, and energy for a compressible fluid
with heat conduction and gravity: 
\begin{eqnarray}
\frac{D\rho }{Dt} &=&-\rho \div\mathbf{v} , \tgg{12.1r} \\
\rho \frac{D\mathbf{v}}{Dt} &=&\div\mathbf{T-}\rho \mathbf{g},
\tgg{12.1v} \\
\rho \frac{D\epsilon }{Dt} &=&\mathbf{T:}\nabla \mathbf{v-}\div%
\mathbf{q} . \tgg{12.1e}
\end{eqnarray}
Here $\mathbf{T}$ is the Cauchy stress tensor, $\rho $ is the density, $%
\mathbf{v}$ is the velocity, $\epsilon $ is the specific internal energy
density, $-\mathbf{g}$ is the gravitational acceleration, $\mathbf{q}$ is
the heat flux, and $D/{Dt}= \partial/{\partial t}+\mathbf{v}\cdot
\nabla .$ 

To account for the influence of density gradients on energy in a
manner compatible with the second law of thermodynamics, we apply a
theory for Korteweg-type fluids described by J.~E. Dunn and J. Serrin
\cite{bib6}. A recent review of related theories and their applications to
diffuse-interface modeling has been given by Anderson, McFadden and
Wheeler \cite{AMW}. Dunn and Serrin replace the energy balance 
equation with
\begin{equation}
\rho \frac{D\epsilon }{Dt}=\mathbf{T:}\nabla \mathbf{v}-\div\mathbf{q}%
+\div\mathbf{u}  \label{12e2}
\end{equation}
where $\div\mathbf{u}$ represents a contribution made to the power by
capillary effects due to the strong density gradients.  They then
postulate that $\epsilon$, $\mathbf{T}$, $\mathbf{q}$, $\mathbf{u}$,
as well as $\psi$, the specific Helmholtz free energy density, and $\ENT$,
the specific entropy density, are given by constitutive relations
that depend only on the pointwise values of $\rho$, $\thett$,
$\nabla\rho$, $\nabla\nabla\rho$, $\nabla\thett$, and $\nabla\mathbf{v}$,
where $\thett$ is the temperature. 

In order to guarantee that the equations of motion are compatible with
the second law of thermodynamics as expressed by the Clausius-Duhem
inequality, Dunn and Serrin deduce that 
the constitutive relations must be such that
\begin{equation}
\psi=\bar\psi(\rho,\thett,|\nabla\rho|^2), \quad
\ENT=-\dd{\bar\psi}{\thett},\quad 
\epsilon=\bar\psi-\thett\dd{\bar\psi}{\thett}, \tgg{12ds1}
\end{equation}
and
\begin{equation}
 \mathbf{u}=-\rho(\nabla \cdot \mathbf{v})m\nabla\rho+\mathbf{\omega },
\tgg{12u}
\end{equation}
where $m=2\rho(\partial\bar\psi/\partial M)(\rho,\thett,M)$ 
with $M=|\nabla\rho|^2$.  
Furthermore, in the case without viscosity, the stress must take the form
\begin{equation}
\mathbf{T}=
\left(-\rho^2\dd{\bar\psi}{\rho}+\rho\div(m\nabla\rho)\right)
\mathbf{1}- m\nabla\rho\otimes\nabla\rho. \tgg{12T1}
\end{equation}
In (\rf{12u}), the quantity $\mathbf{\omega }$ measures the ``static''
part of the capillary work flux $\mathbf{u}$. For a class of materials
including those we shall consider, $\div\mathbf{\omega}=0$, so
$\mathbf{\omega}$ has no effect on the energy balance equation and can
be ignored.

We choose the simplest possible form for the specific Helmholtz free energy
density $\psi$ that is consistent with Dunn and Serrin's
theory. Namely, we suppose that $m$ is constant, and that
\begin{equation}
\psi =\hat{\psi}(\rho ,\thett )+\frac m{2\rho }|\nabla\rho|^2. \tgg{12psi}
\end{equation}
This yields an expression for total free energy that appears in van
der Waals' theory of capillarity \cite{bib18}, for example. From (\rf{12ds1})
it follows that
\begin{equation}
\epsilon=\hat\epsilon(\rho,\thett)+\frac m{2\rho }|\nabla\rho|^2,
\quad \ENT=-\dd{\hat\psi}{\thett}(\rho,\thett),
\tgg{12eps}
\end{equation}
where $\hat\epsilon=\hat\psi-\thett(\partial\hat\psi/\partial\thett)$. 
We assume that
the heat flux is given by $\mathbf{q}=$ $-\kappa \nabla \thett $,
with thermal conductivity coefficient $\kappa =\hat{\kappa}(\rho,\thett)$.
And we suppose that the stress is given by (\rf{12T1}) with the
addition of Newtonian viscosity terms, so that
\begin{equation}
\mathbf{T} =\left(-p+\frac m2|\nabla\rho|^2+m\rho\Delta\rho\right)
\mathbf{1}- m\nabla\rho\otimes\nabla\rho
+\lambda (\div\mathbf{v})\mathbf{1}+2\mu \mathbf{D}, \tgg{12T2}
\end{equation}
where the pressure $p$ is given by 
\begin{equation}
p=\hat p(\rho,\thett)=\rho^2\dd{\hat\psi}{\rho}(\rho,\thett),
\tgg{12phat}
\end{equation}
$\mathbf{D}=\frac 12(\nabla \mathbf{v}+\nabla \mathbf{v}^T)$ is the
symmetric part of $\nabla \mathbf{v}$, and the viscosity coefficients
have the form $\lambda =\hat{\lambda}(\rho ,\thett )$, 
$\mu=\hat{\mu}(\rho ,\thett )$. For compatibility with the Clausius-Duhem
inequality \cite{bib6}, one requires that $\lambda+\frac23\mu\ge0$.

When equations (\rf{12u}), (\rf{12T2}), (\rf{12psi}) and (\rf{12eps})
are substituted into the balance laws (\rf{12.1r}), (\rf{12.1v}),
(\rf{12e2}), the result is the following system of equations for a
viscous, heat-conducting fluid with capillary stresses:
\begin{eqnarray}
 \rho\frac{D\mathbf{v}}{Dt}+\nabla p&=&-\rho \mathbf{g} +m\rho \nabla \Delta
 \rho +\nabla (\lambda \div\mathbf{v})+ \div(2\mu\mathbf{D}),
\tgg{12.2v} \\[4pt]
 \frac{D\thett}{Dt} &=&\left(1-\frac{c_v}{c_p}\right) \frac{K_T}{\alpha_p} 
 \frac{Dp}{Dt}+(\rho c_p)^{-1}\div(\kappa \nabla \thett ) \nonumber\\[4pt]
 &&+(\rho c_p)^{-1} \left(2\mu \mathbf{D}:\mathbf{D}
 +\lambda (\div\mathbf{v})^2 \right),  
\tgg{12.2t}\\[4pt]
 K_{T}\frac{Dp}{Dt} &=&\alpha _p\frac{D\thett }{Dt}-\div\mathbf{v}.  
\tgg{12.2p}
\end{eqnarray}
Here $K_{T}= \rho^{-1} (\partial \rho/\partial p) _\thett $ 
is the isothermal compressibility, 
$\alpha _p= -\rho^{-1} (\partial \rho/\partial \thett) _p $ 
is the isobaric thermal expansion coefficient, and 
\[
c_v=\hat{c}_v(\rho ,\thett )=\dd{\hat\epsilon}{\thett}, \quad 
c_p=\hat{c}_p(\rho ,\thett )=c_v+ \rho^{-1}\alpha_p \thett \dd{\hat p}{\thett}
\]
are the specific heat capacities at constant volume and at constant
pressure, respectively.  The system
(\rf{12.2v})--(\rf{12.2p}) differs from the standard system for a
viscous, heat-conducting fluid by the addition of the single term
$m\rho\nabla\Delta\rho$ in the momentum equation.

\section{Equations for slow flows with weak gravity}\label{WeakG}
\setcounter{equation}{0}

\subsection{Scaling for the flow regime}\label{FlowRegime}

In order to see how the system of equations (\rf{12.2v})--(\rf{12.2p}) can be
appropriately simplified, we non-dimensionalize and scale the variables in a
manner appropriate to a typical experiment, as follows. The critical
density, critical pressure, and critical temperature are denoted $\rho _c,$
$p_c,$ and $\thett _c$, respectively. Let $x_a$ be a characteristic length,
$t_a$ a characteristic time, and $v_a=x_a/t_a$. Let $c_{p_a},$ $c_{v_a}$ be
characteristic specific heat capacities at constant pressure and at
constant volume respectively, and let $\Gamma =c_{p_a}/c_{v_a}$. Let
$\kappa _a$ be a characteristic thermal conductivity, and let
$\lambda_a$, $\mu_a$ be characteristic viscosity coefficients (for
convenience we assume $\lambda_a=\mu_a$). Let $g_a=x_a/t_a^2$, and
introduce the following non-dimensional variables and constants
(non-dimensional quantities are indicated by an asterisk, $^{*}$ ):
\begin{eqnarray*}
& \rho =\rho_c\rho ^{*},\quad p=p_c(1+p_ap^{*}),
\quad \thett =\thett_c(1+\thett _a\thett ^{*}), &\\
&\mathbf{x} =x_a\mathbf{x}^{*},\quad t=t_at^{*},
\quad \mathbf{v}=v_a\mathbf{v}^{*},\quad \mathbf{g}=g_a\mathbf{g}^{*},  &\\
&c_v =c_{v_a}c_v^{*},\quad c_p=c_{p_a}c_p^{*},
\quad \kappa =\kappa_a\kappa ^{*},\quad \lambda =\lambda_a\lambda ^{*},
\quad \mu =\mu _a\mu^{*}, &\\
&\displaystyle
K_{T}^{*} =\frac 1{\rho ^{*}}\left( \frac{\partial \rho ^{*}}
{\partial p^{*}}\right) _{\thett ^{*}},
\quad \alpha _p^{*}= -\frac 1{\rho^{*}}\left( \frac{\partial \rho ^{*}}
{\partial \thett ^{*}}\right) _{p^{*}}. &
\end{eqnarray*}
Here, $p_a$ and $\thett _a$ are non-dimensional scales characterizing the
deviation of the pressure and temperature from critical. One tries to choose
the scales so that the dimensionless variables $p^{*},$ $\thett ^{*},\mathbf{%
v}^{*}$ and their derivatives are of the order of 1 in the flow regime of
interest.

The non-dimensional equations corresponding to equations
(\rf{12.2v})--(\rf{12.2p}) for a viscous fluid are 
\begin{eqnarray}
\rho ^{*}\frac{D\mathbf{v}^{*}}{Dt^{*}}&=&-M^{-2}\nabla p^{*} +c\rho
^{*}\nabla \Delta \rho ^{*}-\rho ^{*}\mathbf{g}^{*} 
\nonumber \\
&&+\text{Re}^{-1}
\left(\nabla (\lambda ^{*}\div 
\mathbf{v}^{*})  +
\nabla \cdot (2\mu ^{*}\mathbf{D}^*) \right),  \tgg{13.1v} \\
\frac{D\thett ^{*}}{Dt^{*}} &=&\left( 1-\Gamma ^{-1}\frac{c_v^{*}}{c_p^{*}}%
\right) 
\frac{K_T^{*}}{\alpha _p^{*}} \,
\frac{Dp^{*}}{Dt^{*}}+\frac{D_T}{\rho^{*}c_p^{*}}
\div(\kappa ^{*}\nabla \thett ^{*})  \nonumber \\
&&+S\left(\frac{2\mu ^{*}}{\rho ^{*}c_p^{*}} \mathbf{D}^*:\mathbf{D}^*
+\frac{\lambda ^{*}}{\rho ^{*}c_p^{*}}
(\div\mathbf{v}^{*})^2  \right),
\tgg{13.1t}\\
K_{T}^{*}\frac{Dp^{*}}{Dt^{*}} &=&\alpha _p^{*}\frac{D\thett ^{*}}{%
Dt^{*}}-\div\mathbf{v}^{*}.  \tgg{13.1p}
\end{eqnarray}
where the gradient and the divergence are taken with respect to the
non-dimensional spatial variable, and $D/{Dt^{*}}= 
\partial/{\partial t^{*}}+\mathbf{v}^{*}\cdot \nabla .$

The dimensionless constants $M^2$, $D_T$, $c$, Re$^{-1}$, 
and $S$ are defined by 
\begin{eqnarray*}
&\displaystyle
 M^2 =\frac{\rho _cv_a^2}{p_cp_a},\quad D_T=\frac{\kappa _at_a}{\rho
_cc_{p_a}x_a^2},\quad c=\frac{m\rho _ct_a^2}{x_a^4}, 
\quad 
 \text{Re}^{-1} =\frac{\mu _at_a}{\rho _cx_a^2},\quad 
S=\frac{\mu_a}{\rho_c  c_{p_a}\thett _c\thett _a t_a}. &
\end{eqnarray*}
The parameter $M$ is proportional to the Mach number $v_a/c_s$, where
the sound speed is given by $c_s^2=(\partial p/\partial\rho)_s=
c_p/c_v\rho K_T$.
Re is the Reynolds number, $D_T$ is a non-dimensional diffusivity,
and $c$ is a non-dimensional coefficient of capillarity.

A typical flow regime in which we are interested is one considered in the
paper by Boukari, Pego, and Gammon \cite{bib3}. The fluid is xenon,
with critical parameters
\[
\rho _c=1.11\times 10^3\frac{\text{kg}}{\text{m}^3}, \quad
p_c=5.84\times 10^6 \text{ Pa}, \quad \thett _c=289.72 \text{ K}.
\]
A typical experimental cell is about half a centimeter in radius, so we
choose $x_a=10^{-3}$ m. The time scale of interest ranges from a
fraction of a second to hours. For now we take $t_a=1$~s and postpone
further discussion. The effect of earth's gravity tends to become
important within about 30 mK of $\thett _c$, so we take
$\thett_a=10^{-4}$. In this temperature range and near the critical density,
an appropriate model equation of state is the restricted cubic model,
with coefficients for xenon (see \cite{bib12} and the references therein).
From this model, as in \cite{bib3} we find it appropriate to take
\[
p_a =6\thett_a,
\quad c_{p_a}\approx 3.3\times 10^6
\frac{\text{J}}{\text{kg}\cdot \text{K}},
\quad \Gamma ^{-1} \approx 1.8\times 10^{-4}.
\]
We have not introduced a separate scale for deviations of the density
from critical because these can be rather large, of the order of
$10\%$. 

We estimate a characteristic value for thermal conductivity as in \cite{bib3},
using the approach described in \cite{B27,B28} to calculate the
background term and divergent part near the critical point.  
To estimate the viscosity, we use the results of 
\cite[Table III]{B27}, also see \cite{bib16}.
As a result we find it appropriate to take
\[
\quad \kappa _a\approx 2.5\times 10^{-1}
\frac{\text{J}}{\text{s}\cdot \text{m}\cdot \text{K}},
\quad \mu_a\approx 5\times 10^{-5}\text{ Pa}\cdot \text{s}.
\]

Finally, we estimate the constant capillarity coefficient $m$ using the
power law representation $m\sim \xi^2/\chi_T$ derived by Rowlinson and
Widom \cite{bib18}, where the correlation length $\xi$ and susceptibility
$\chi_T=\rho ^2K_T$ at the critical density are given \cite{bib19} by the power laws
\begin{equation}
\xi \sim \xi _0| \Delta \thett ^{*}| ^{-\nu } ,\quad
\chi _T \sim \frac{(\rho _c)^2C}{p_c}| \Delta \thett ^{*}|^{-\gamma }.
\label{13.powers}
\end{equation}
Here
$\Delta \thett^{*} =(\thett -\thett _c)/\thett_c=\thett _a\thett ^{*}
\approx 10^{-4}$, and we use from \cite{bib19} the critical exponents
$\gamma =1.19$ and $\nu=0.63$, and for xenon in the range  $\thett>\thett_c$
approximately $\xi _0=1.9\times 10^{-10},$ $C=.0813$. Using these values
we obtain the estimate
\[
m \approx \frac{(5.84\times 10^6)(1.9\times
10^{-10})^2}{(1.11\times 10^3)^2(.0813)}(10^{-4})^{-.07}
\approx 4\times 10^{-18}.
\]

From the estimates above, we obtain the
following estimates for the dimensionless parameters: 
\begin{eqnarray*}
&M^2\approx 3\times 10^{-7},\quad D_T \approx6\times 10^{-5},\quad 
c\approx 4.4\times 10^{-3},
\quad &\\ & 
\text{Re}^{-1}\approx 4.5\times 10^{-2},\quad S \approx 5\times 10^{-13},&
\end{eqnarray*}
and we find
\[
|\mathbf{g}^{*}|=9.81\times 10^3,
\quad \frac{c_p^{*}}{c_v^{*}}\sim 1,
\quad \left( \frac{\partial p^{*}}{\partial\thett ^{*}}
\right) _{\rho ^{*}}  \sim 1.
\]
The sound speed $c_s\approx 80$ m/s.

Regarding these parameters, several points are worthy of comment.
First, note the effect of considering longer time scales. As $t_a$
increases, $S$ decreases and $M^2$ decreases quadratically, $D_T$ and
Re${}^{-1}$ increase, and $c$ and $\mathbf{g}^*$ increase quadratically. 
Second,
experiments performed in low earth orbit are reported to experience
typical accelerations of $10^{-4}$ to $10^{-6}$ times earth's gravity
\cite{bib12}; this would make $\mathbf{g}^*$ of order 1.  
(Another way to obtain $\mathbf{g}^*$ of order 1 is to consider a 
faster time scale like $t_a=.01$ s.)
 Also, we note
that the value of $c$ becomes of order 1 when the characteristic
length is replaced by a capillary length $x_{\text{cap}}$ for which
$1={t_a^2m\rho _c}/{x_{\text{cap}}^4}.$ 
For our flow regime, we estimate $x_{\text{cap}}\approx x_a(4.4\times
10^{-3})^{1/4}\approx 260$ microns.
In the recent ZENO experiment, observations were
performed using light scattering through a fluid layer 100 microns
thick \cite{Zeno}.

We should also comment on the effect of proximity to the critical temperature. 
As $\thett_a$ approaches zero, we have seen that the capillary
coefficient diverges very weakly, with exponent $\gamma-2\nu\approx-0.07$. 
The viscosity and the sound speed also diverge at a very slow rate,
changing only modestly over the experimental range of interest.
The specific heat $c_v$ also diverges weakly, with exponent
$-\alpha\approx-0.11$. The compressibility, thermal expansivity and specific
heat $c_p=c_v+ \chi_T\thett\rho^{-3}(\partial\hat p/\partial\thett)^2$
diverge strongly,
all with exponent $-\gamma\approx-1.19$. The thermal conductivity diverges
less strongly, like $c_p/\xi\mu$ \cite[p.~22]{bib19}, with
approximately the exponent
$-(\gamma-\nu)\approx-0.56$.  So we see that as $\thett_a$ approaches zero,
none of the nondimensional constants above has a very strong dependence on
$\thett_a$, though we can expect the nondimensional diffusivity $D_T$
and $S$ to decrease.

\subsection{Reduced equations for slow flows}\label{Reduce}

To obtain simplified equations for compressible flows, we proceed
formally in a manner motivated by the treatments of Rehm and Baum \cite{bib17}
and Majda and Sethian \cite{bib11}. Since we are interested in longer time
scales, we regard $M^2$ and $S$ as small and let
$p_0$, $\thett_0$, $\mathbf{v}_0$ denote assumed asymptotic limits of
$p^*$, $\thett^*$, $\mathbf{v}^*$ respectively, as $M^2$ and $S$ are
taken to zero. In this process we regard $\mathbf{g}^*$ as fixed and 
of order one, corresponding to a low-gravity environment for our flow 
regime.
Multiplying the momentum equation (\rf{13.1v}) by $M^2$ 
and taking $M^2$ to 0, we require
\begin{equation}
\nabla p_0(\mathbf{x}^{*},t^{*}) =0,  \tgg{14grad}
\end{equation}
and therefore $p_0=p_0(t^*)$ is constant in space.
From (\rf{13.1t})--(\rf{13.1p}), the asymptotic equations for 
temperature and pressure are
\begin{eqnarray}
\frac{D\thett _0}{Dt^{*}}&=&
\left( 1-\Gamma ^{-1}\frac{c_{v0}}{c_{p0}}\right)
\frac{K_{T0}}{\alpha _{p0}}\frac{dp_0}{dt^{*}} +
\frac{D_T}{\rho_0c_{p0}}\div(\kappa _0\nabla \thett _0),  \tgg{14t}\\
K_{T0}\frac{dp_0}{dt^{*}}&=&\alpha _{p0}\frac{D\thett _0}{Dt^{*}}-
\div\mathbf{v}_0  \tgg{14p0} .
\end{eqnarray}
Here the asymptotic density $\rho_0(\mathbf{x}^{*},t^{*})=
\hat\rho^*(p_0(t^*),\thett_0(\mathbf{x}^{*},t^{*}))$ from the equation of
state, and elsewhere the subscript 0 indicates a nondimensional
coefficient that is evaluated at $(\rho _0,\thett _0)$, 
e.g., $\kappa _0(\mathbf{x}^{*},t^{*})=\kappa ^{*}(\rho _0,\thett _0)$.
Equation (\rf{14p0}) implies the mass continuity equation
$D\rho _0/Dt^{*}+\rho _0\div\mathbf{v}_0=0$.

Next we derive a reduced momentum equation. We suppose the flow
occurs in a bounded domain $\Omega$ with $\mathbf{v}=0$ on the boundary
$\partial\Omega$. We use the fact that every square-integrable vector
field $\mathbf{v}$ has a unique orthogonal decomposition of the form
\begin{equation}
\mathbf{v}=\mathbf{w}+\nabla\phi, \quad\text{where}\quad
\div\mathbf{w}=0 \text{  and  }
\left.\mathbf{w}\cdot\mathbf{n}\right|_{\partial\Omega}=0
\label{15vwphi}
\end{equation}
($\mathbf{n}$ is the outward unit normal to $\partial\Omega$).
We write $P\mathbf{v}=\mathbf{w}$, so $P$ denotes the orthogonal
projection of square-integrable vector fields onto solenoidal vector
fields. Observe that if $P\mathbf{f}=0$, then
$\mathbf{f}=\nabla h$ for some function $h$, and conversely.
Applying $P$ to the momentum equation (\rf{13.1v}), then, eliminates 
the term of order $M^{-2}$. Taking $M^2$ to 0 produces
\[
0= P\left[-\rho _0\frac{D\mathbf{v}_0}{Dt^{*}}
-\rho _0\mathbf{g}^{*} 
+c\rho _0\nabla \Delta \rho _0
+\frac1{\text{Re}}\left(
\div(\mu _0(\nabla \mathbf{v}_0+\nabla \mathbf{v}_0^{T}))
+\nabla (\lambda _0\div\mathbf{v}_0) \right)
\right]  
\]
where $\mu _0=\mu ^{*}(\rho_0,\thett _0)$,
$\lambda _0=\lambda ^{*}(\rho_0,\thett_0).$

From this we infer that there must be a scalar function
$\pnon=\pnon(\mathbf{x}^{*},t^{*})$ such that the expression above in brackets
equals $\nabla\pnon$, that is,
\begin{eqnarray}
\rho _0\frac{D\mathbf{v}_0}{Dt^{*}}+\nabla\pnon &=&
-\rho _0\mathbf{g}^{*} +c\rho _0\nabla \Delta\rho _0
+\frac1{\text{Re}} \left(
\div(\mu_0(\nabla\mathbf{v}_0+\nabla \mathbf{v}_0^{T}))
+ \nabla (\lambda_0\div\mathbf{v}_0)\right). \phantom{mmmm}
\tgg{14v}
\end{eqnarray}
Note that at this point we do not presume that the pressure $p^*\approx
p_0+M^2\pnon$ to order $M^2$. Klainerman and Majda \cite{bib8} have found
that at order $M^2$ there is an acoustic correction to pressure that
depends on fast time and space scales.

Equations (\rf{14t}), (\rf{14p0}), and (\rf{14v}) are the simplified
equations for a viscous fluid.  In order that our approximations be
self-consistent, we must require that at the initial time, as $M^2$
and $S$ tend to zero we have
\[
p^{*}(\mathbf{x}^{*},0)\to p_0(0),\quad
\thett ^{*}(\mathbf{x}^{*},0) \to\thett _0(\mathbf{x}^{*},0), \quad
\mathbf{v}^{*}(\mathbf{x}^{*},0)\to\mathbf{v}_0(\mathbf{x}^{*},0).
\]

\subsection{Reformulation}\label{Reform}

The full system (\rf{12.2v})--(\rf{12.2p}) is appropriate for
describing compressible fluid flow on acoustic time scales. In the
flow regime for xenon that we have described, such time scales are
short, since the sound speed is of the order of tens of meters per
second.  The simplified system (\rf{14t}), (\rf{14p0}),
(\rf{14v}) represents an `acoustic filtering' of the full system that
describes flow on time scales that are long compared to acoustic. The
pressure is maintained spatially constant through a process mediated
by sound wave propagation. (An asymptotic description of this process
for a near-critical van der Waals fluid was given by Zappoli and Carles
\cite{ZC} in one dimension with no viscosity.)

As is easy to check, both the full system and the simplified system are
compatible with the Clausius-Duhem inequality
\[
\rho \frac{D\ENT }{Dt}+\div\left( \frac{\mathbf{q}}\thett \right) \geq
0. 
\]

We next reformulate the simplified system (\rf{14t}), (\rf{14p0}),
(\rf{14v}) into an equivalent form which better reveals its evolutionary
character. This reformulation will be necessary for our future purpose
of analyzing the initial-value problem. We will omit the subscripts and
superscripts and return to dimensional quantities for notational
convenience, writing
\[
 \rho =\rho_c\rho_0,\quad \pzero=p_c(1+p_ap_0),
\quad \thett =\thett_c(1+\thett _a\thett_0), \quad
\quad \mathbf{v}=v_a\mathbf{v}_0.
\]

We employ the decomposition $\mathbf{v=w}+\nabla\phi$ described in
(\rf{15vwphi}). Substituting the temperature equation (\rf{14t})
into the pressure equation (\rf{14p0}) and solving for
$\div\mathbf{v}$, we obtain
\begin{equation}
\div\mathbf{v}=\Delta \phi =
-\frac{c_vK_T}{c_p}\frac{d\pzero}{dt}+
\frac{\alpha _p}{\rho c_p}\div(\kappa \nabla \thett ).  \tgg{15divv}
\end{equation}
This elliptic equation has a solution with
$\nabla\phi\cdot\mathbf{n}=\mathbf{v}\cdot\mathbf{n}=0$
on $\partial\Omega$
(and then $\nabla\phi$ is uniquely
determined), if and only if a solvability condition holds, namely
\begin{equation}
\int_\Omega\left( -\frac{c_vK_T}{c_p}\frac{d\pzero}{dt}+
\frac{\alpha _p}{\rho c_p}\div(\kappa \nabla \thett )\right)dx=0.  \tgg{15solv}
\end{equation}
Solving this equation for ${d\pzero}/{dt}$, we get 
\begin{equation}
\frac{d\pzero}{dt}=H(t):=\frac{\int_\Omega \left(\alpha _p/\rho c_p)
\div(\kappa \nabla \thett \right)dx}{\int_\Omega (c_vK_T/c_p)\,dx} .
\tgg{15p}
\end{equation}
This integro-differential equation is the pressure evolution equation.
The remaining equations of the system can be written as
\begin{eqnarray}
\frac{D\thett }{Dt} &=&\left( 1-\frac{c_v}{c_p}\right)
 \frac{K_T}{\alpha_p}H(t)
 +\frac{1}{\rho c_p}\div(\kappa \nabla \thett ),  \tgg{15t}
\\
\rho \frac{D\mathbf{w}}{Dt} &=&-\nabla \pdim
 -\rho\mathbf{g}
+c\rho \nabla \Delta \rho
-\rho \frac{D(\nabla \phi )}{Dt}  \nonumber\\
 &&+\div\left(\mu (\nabla (\mathbf{w}+\nabla \phi )+
 \nabla (\mathbf{w}+ \nabla \phi )^{T})\right)+
 \nabla (\lambda \Delta \phi ),  \tgg{15w}
\\
\Delta \phi &=&-\frac{c_vK_T}{c_p}H(t)+\frac{\alpha_p}{\rho c_p}
 \div(\kappa \nabla \thett ),  \tgg{15phi}
\\
\div\mathbf{w}&=&0, \tgg{15divw}
\end{eqnarray}
along with the equation of state $\rho=\hat{\rho}(\pzero,\thett )$.
(Here $\pdim=p_cp_aM^2\pnon$.)

From (\rf{15phi}) evaluated at time $t=0$, we obtain the compatibility
condition
\begin{equation}
\Delta\phi|_{t=0}
=\left.\left[-\frac{c_vK_T}{c_p}H(t)+
\frac{\alpha _p}{\rho c_p}
\div(\kappa \nabla \thett )\right]\right|_{t=0} . \tgg{15phi0}
\end{equation}
This equation imposes a constraint on the gradient part
$\nabla\phi(\mathbf{x},0)$ of the initial velocity field
$\mathbf{v}(\mathbf{x},0)$, and leaves freedom for the choice of the
initial solenoidal component $\mathbf{w}(\mathbf{x},0)$, as long as
$\div\mathbf{w}(\mathbf{x},0)=0.$

As we show in \cite{DP2}, to solve the initial-value problem for these
equations, it is appropriate to specify initial data for the
leading-order pressure, temperature, and divergence-free part of the
velocity field:
\begin{equation}
\pzero(0)=\pzero_0,\quad \thett(\mathbf{x},0)=\thett_0(\mathbf{x}),\quad
\mathbf{w}(\mathbf{x},0)=\mathbf{w}_0(\mathbf{x})
\tgg{15init}
\end{equation}
where  $\div\mathbf{w_0}=0$.
Initial data for the velocity will take the form
\begin{equation}
\mathbf{v}(\mathbf{x},0) = \mathbf{v}_0(\mathbf{x}) =
\mathbf{w}_0(\mathbf{x})+\nabla\phi_0(\mathbf{x}),
\label{15.v0}
\end{equation}
where $\nabla\phi_0$ is determined by the compatibility condition
(\rf{15phi0}).

\subsection{The adiabatic time scale}

We conclude this section by indicating how the simplified
equations (\rf{15p})--(\rf{15divw}), in nondimensional form, can be
used to very roughly estimate the time scale $t_1$ for the adiabatic
mechanism described in the introduction to produce a rapid bulk
temperature response to boundary heating.  

The nondimensional form of
(\ref{15p}) is 
\begin{equation}
\frac{dp_0}{dt^*}=\frac{ \Gamma D_T\int_\Omega (\alpha_{p0}/\rho_0 c_{p0})
\nabla \cdot(\kappa_0 \nabla \thett_0 ))dx}
{\int_\Omega (c_{v0}K_{T0}/c_{p0})\,dx} .
\tgg{15p0}
\end{equation}
We consider a homogeneous fluid initially at equilibrium, whose
boundary temperature is raised rapidly.  As the boundary temperature
is changed, a thin thermal boundary layer is created next to the
wall. The width of this layer increases with time through heat
diffusion, so is roughly given by $\sqrt{t^*D_T}$ using (\rf{14t}).  In
the boundary layer, we may roughly approximate the nondimensional
temperature $\thett_0$ by a function of the form $f(s/\sqrt{t^*D_T})$,
where $s$ is the distance to the boundary. Treating the coefficients
in (\rf{15p0}) as constant, we estimate $\Delta\thett_0\approx
(t^*D_T)^{-1}f''(s/\sqrt{t^*D_T})$ in the boundary layer. The integrand
in the numerator is then of order $(t^*D_T)^{-1}$ in the boundary
layer and zero elsewhere, and the integrand in the denominator is of
order 1. Suppose the fluid domain is a cube with dimensional side
length $L=x_aL^*$. Then the order of $dp_0/dt^*$ is given by

\begin{equation}
\frac{dp_0}{dt^*}\sim \Gamma D_T\frac{6L^{*2}\sqrt{t^*D_T}(t^*D_T)^{-1}}
{L^{*3}} = \left(\frac{36\Gamma^2D_T}{L^{*2}t^*}\right)^{1/2}.
\tgg{15pest}
\end{equation}
The time integral of 
this expression produces an order one change in $p_0$ (hence in
$\thett_0$), when $t=t_1=t^*t_a=L^2\Gamma^{-2}/144D$, where 
$D=\kappa_a/\rho_cc_{p_a}$ is the characteristic thermal diffusivity.
For the fluid parameters corresponding to the flow regime which
we have described above, in a cell with side length $L=10^{-2}$ m we estimate 
$t_1\approx 3.3\times10^{-4}$~s. 

This time is longer but not much longer than the acoustic time $t_c$ for a
sound wave to cross the cell, given by $t_c=L/c_s\approx
1.25\times10^{-4}$ s.  It is unrealistic to expect, however, that the
boundary temperature can be raised so rapidly in experiment.  So what
this estimate indicates is that for processes in which the boundary
temperature varies slowly compared to the acoustic time, the
boundary-to-bulk coupling provided by the integro-differential
equation for the mean pressure is efficient in effecting bulk
temperature changes.

The above estimate for $t_1$ is consistent with the results of Onuki and
Ferrell \cite{bib14}, except for the geometric factor of $144$ appearing in the
denominator. We expect that diffusion dominates the equilibration at 
approximately the time $t_d={L^2}/{144D}\approx 10^4$~s.
Onuki, Hao and Ferrell \cite{bib15} characterize the intermediate regime
between the long times $t/{t_1}\geq \Gamma^2$ and the short times
$t/{t_1}=O(1)$ by the geometrical mean
${t_{\text{int}}}/{t_1}\equiv\Gamma$ so that
$t_{\text{int}}=\Gamma t_1$. For our flow regime we estimate
$t_{\text{int}}\approx 2$~s.

\section{Multi-dimensional flows with strong gravity}\label{StrongG}
\setcounter{equation}{0}

\subsection{Motivation}

In this section, we re-examine the equations of motion in the case of
strong gravity. Motivating us is the problem of describing
near-critical fluid flows and equilibration in earth's gravity.
Recall that $|\mathbf{g}^*|\sim 10^4$ in the flow regime considered in
section \ref{FlowRegime} with earth's gravity.
The simplified system (\rf{15p})--(\rf{15divw}) fails to capture some
key features of equilibration in this situation.

In equilibrium, temperature is constant and density is stratified
according to the basic equation of hydrostatic balance,
\begin{equation}
\nabla p = -\rho \mathbf{g},
\tgg{31.hydro}
\end{equation}
and the equation of state.
(We will neglect capillarity in most of this section.)
Denoting equilibrium temperature by $\theteq$ and density by
$\rhoeq(z)$, the equilibrium density gradient satisfies
\[
\frac{d\rhoeq}{dz}(z)= -\chi_T(\rhoeq(z),\theteq)g.
\]
As temperature
approaches $\thett_c$, the critical temperature,
$\chi_T(\rho_c,\theteq)$, the susceptibility on the critical isochore,
diverges as in (\ref{13.powers}). Thus the density gradient develops a
singularity at the level of critical density, and the density profile
becomes highly nonlinear. In Fig.\ 1 we plot density profiles
for xenon in equilibrium at 1G, using the restricted cubic 
equation of state as in \cite{bib3}.

\begin{figure}
\centering
\epsfig{file=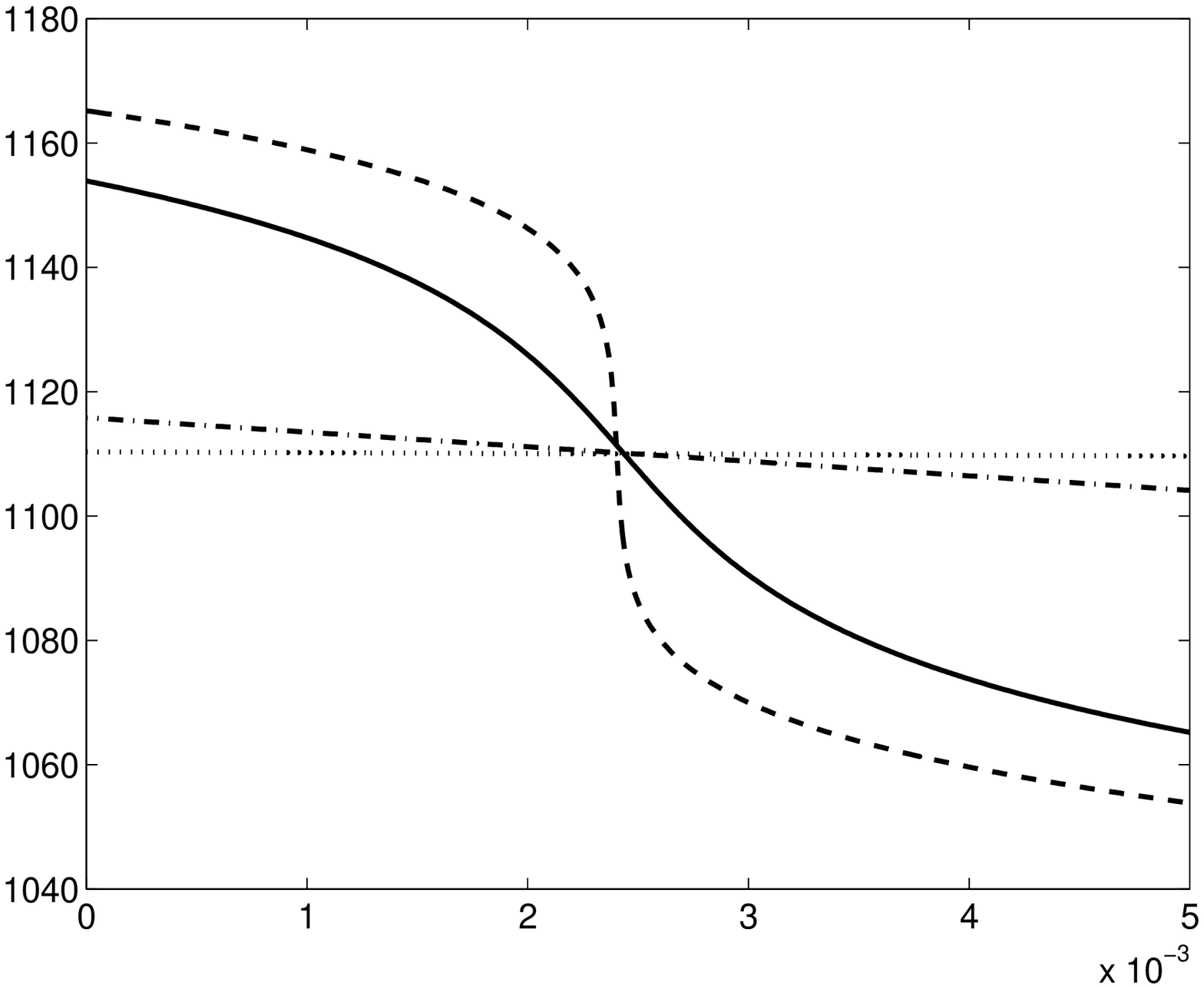,height=3.5in} 

\vspace{10pt}
Figure 1: Density vs height for near-critical xenon at 1G.  \hfil\linebreak
$\thett-\thett_c$ in mK: 1 (dash), 10 (solid), 100 (dash-dot), 1000 (dot)
\end{figure}

For the simplified system (\rf{15p})--(\rf{15divw}), however, the
density in equilibrium is constant, given by
$\rho=\hat\rho(\peq,\theteq)$. The hydrostatic balance
$\nabla\pdim=-\rho\mathbf{g}$ from (\rf{15w}) can be interpreted as 
supplying an $O(M^2)$ correction to the leading order pressure.
This correction is linear in
$z$ and could be used to generate a density correction (by linearizing
the equation of state, for example).  But as it stands, the system 
(\rf{15p})--(\rf{15divw}) relies only on the leading-order density to
determine the fluid flow. One can therefore expect this system to be
inaccurate in modeling phenomena such as deep convection and gravity
waves in near-equilibrium states. 

In order to model related phenomena, 
researchers studying small-scale atmospheric circulations frequently
approximate the continuity equation with the `anelastic' continuity equation
\begin{equation}
\nabla \cdot (\bar{\rho}\mathbf{v})=0,  
\label{eqn1}
\end{equation}
where $\bar{\rho}(z)$ is usually defined either as the density in an
adiabatically stratified, horizontally uniform reference state, or as
the horizontally-averaged actual density.  Batchelor \cite{B} introduced an
equation equivalent to (\ref{eqn1}). The name
`anelastic' was given by Ogura and Phillips \cite{OP}, who derived
(\ref{eqn1}), together with approximate momentum and thermodynamic
equations, through a systematic scale analysis.  Important assumptions
in their analysis are that: (i) all deviations $\delta \Tpot$
of the `potential temperature' from some constant mean value $\Tpot _a$
are small
(this is equivalent to a similar statement for entropy variations),
and (ii) the time scale of the disturbance is similar to the time
scale for gravity wave oscillations. The terms neglected in their
approximation are formally an order $\epsilon =\delta \Tpot /\Tpot _a$
smaller than those which are retained.  Their anelastic system does
not support sound waves, does support gravity waves and conserves
energy.  Ogura and Phillips define $\bar{\rho}$ as the density in an
adiabatically stratified, horizontally uniform reference state. 

Some of the largest errors in the Ogura-Phillips anelastic
approximation are reported to be generated by large deviations of the
mean state potential temperature (or entropy) from a constant
reference value.
Several authors (Dutton and Fichtl \cite{DF}, Wilhelmson and
Ogura \cite{WO}, Lipps and Hemler \cite{LH}, Durran \cite{D}) have
presented alternative `sound-proof' equations in which thermodynamic
variables associated with the adiabatic reference state are replaced
with horizontally-uniform averaged or approximate actual values.
These authors make different approximations in the momentum equations,
but (except for Durran) they all obtain a continuity equation of the
form (\ref{eqn1}) in which $\bar{\rho}$ is defined as a
horizontally-averaged approximation to the actual density.  

In the remainder of this section, we describe three possible models
for multi-dimensional flows under strong gravity, making
different assumptions about how to balance and approximate terms in
the nondimensionalized system (\rf{13.1v})--(\rf{13.1p}):
\begin{itemize}
\item[(1)] At first we
scale so as to enforce hydrostatic balance at leading order. We find
then that to be consistent, at leading order the thermodynamic
variables must remain horizontally uniform. Also the vertical velocity
must remain horizontally uniform, {\it unless} we assume the entropy is
spatially {\it constant} to leading order.
\item[(2)]  If we indeed assume that entropy is spatially constant to
leading order and also
neglect heat conduction, we obtain a generalization of the anelastic
approximation valid for a general equation of state.  
\item[(3)]
Neither choice so far admits both thermodynamic equilibrium states and
non-stratified multidimensional flows. We find, however, that if we return
to the scaling adopted in section \ref{FlowRegime}, and modify the
momentum equation so that the gravity force term incorporates a
pressure correction self-consistently, then we get a system of the
same formal accuracy that admits physically correct equilibrium states
exactly. Moreover the new system supports gravity waves but not
acoustic waves. 
\end{itemize}

\subsection{Strongly stratified flow}

Because equilibria are governed by the equation of hydrostatic
balance, if gravity is strong it is natural to try and balance the
pressure gradient term with the gravity term in the nondimensional
momentum equation (\ref{13.1v}). Thus we regard $M^{-2}$ and
$|\mathbf{g}^*|$ to be of the same order. This same approach was taken
in \cite{bib3} for one-dimensional flows. 

In this approximation, we rescale the gravitational acceleration, 
writing  $\mathbf{g}^{*}=M^{-2}\mathbf{\bar{g}}$ where 
$\mathbf{\bar{g}}=O(1)$. It will prove 
instructive to replace $(p,\thett)$ by $(p,s)$ as the
thermodynamic state variables, where $s$ is the specific entropy density. 
In terms of these variables, 
we write the equation of state as $\rho=\tilde\rho(p,s)$.
In nondimensional form, we write $\rho^*=\tilde\rho^*(p^*,s^*)$, where
entropy is nondimensionalized via the relation $s=s_as^*$. 
It is convenient to take $s_a=c_{pa}\thett_a$; this will be discussed 
further in the next subsection. 

Starting from the nondimensionalized system
(\ref{13.1v})--(\ref{13.1p}) with the replacement 
$\mathbf{g}^{*}=M^{-2}\mathbf{\bar{g}}$, we proceed as in section
\ref{Reduce} and regard $M^2$ and $S$ as small, obtaining the 
temperature evolution equation (\ref{14t}) and 
continuity equation (\ref{14p0}). In terms of the leading order
pressure $p_0$ and entropy $s_0$, these equations are equivalent to 
\begin{eqnarray}
\rho_0 (1+\thett_a\thett_0) \frac{Ds_0}{Dt^*}&=&
D_T \nabla \cdot (\kappa_0 \nabla \thett_0 ),
\label{32.s}\\[4pt]
 \frac1{\rho_0 c_{s0}^2}\frac{Dp_0}{Dt^*} &=&
\frac{\alpha _{p0}}{\rho_0 c_{p0}}
\left(\rho_0(1+\thett_a\thett_0)\frac{Ds_0}{Dt^*}\right)-\div\mathbf{v}_0.
\label{32.p}
\end{eqnarray}

The leading-order terms in the momentum equation now just yield the
equation of hydrostatic balance, 
\begin{equation}
\nabla p_0 = -\rho_0 \mathbf{\bar{g}} .
\label{32.hydro}
\end{equation}
This equation imposes tight restrictions on the spatial dependence of the
leading-order
thermodynamic state variables. In particular, it is necessary that $p_0$
is a function only of height $z^*$ and time $t^*$ and is independent of
the horizontal variables $(x^*,y^*)$. Then from
(\ref{32.hydro}) it follows that $\rho_0=\rho_0(z^*,t^*)$ is also 
horizontally uniform.
The equation of state now implies that $s_0=s_0(z^*,t^*)$ as well. The entropy
and pressure equations (\ref{32.s})--(\ref{32.p}) now read
\begin{eqnarray}
\rho_0(1+\thett_a\thett_0)
\left(\frac{\partial s_0}{\partial t^*}+w_0 \dd{s_0}{z^*}\right) &=&
D_T\dd{}{z^*}\left( \kappa_0 \dd{\thett_0}{z^*} \right).
\label{32.dsdz} \\[4pt]
 \frac1{\rho_0 c_{s0}^2}\left(\frac{\partial p_0}{\partial t^*}+w_0
\dd{p_0}{z^*}\right)
 &=& D_T\frac{\alpha _{p0}}{\rho_0 c_{p0}}
\dd{}{z^*}\left(\kappa_0 \dd{\thett_0}{z^*} 
\right)-\div\mathbf{v}_0.  
\label{32.pz}
\end{eqnarray}
Here $w_0$ is the vertical component of the velocity $\mathbf{v}_0$.

{\it If we assume} $\partial s_0/\partial z^*\ne0$, then to satisfy
(\ref{32.dsdz}) consistently, {\it the vertical velocity must be
horizontally uniform}: $w_0=w_0(z^*,t^*)$. We can eliminate the horizontal
components of velocity from (\ref{32.pz}) by integrating over $(x^*,y^*)$;
if the cell walls are vertical or periodic we can express the result
in terms of density as 
the usual one-dimensional continuity equation,
\begin{equation}
\frac{\partial \rho_0}{\partial t^*}+\dd{(\rho_0 w_0)}{z^*} = 0.
\label{32.rhoz}
\end{equation}

The system consisting of the three equations (\ref{32.hydro}),
(\ref{32.dsdz}), (\ref{32.rhoz}) is equivalent to the one-dimensional
system (14)--(16) in \cite{bib3} for temperature and pressure, which
was expressed using a Lagrangian variable 
$z'=\int_0^{z^*} \rho_0(h,t^*)\,dh$ in
place of the height $z^*\in[0,L]$. 
With $t'=t^*$, in present notation this system takes the form
\begin{eqnarray}
p_0(z',t')&=& p_a(t')-\bar{g}z' , \\[4pt]
\label{B14}
\dd{\thett_0}{t'}&=&\left(1-\frac{c_{v0}}{\Gamma c_{p0}}\right) 
\frac{K_{T0}}{\alpha_{p0}} 
\frac{dp_a}{dt'} + \frac{D_T}{c_{p0}} \dd{}{z'}
\left(\rho_0\kappa_0 \dd{\thett_0}{z'}\right), \\[4pt]
\label{B15}
\frac{dp_a}{dt'} &=& \frac{ \int_0^m
\rho_0^{-1}\alpha_{p_0}(\partial\thett_0/\partial t')\,dz' }
{\int_0^m \rho_0^{-1}K_{T0}\,dz'} .
\label{B16}
\end{eqnarray}
(Here $m=\int_0^L\rho_0(h,t^*)\,dh$ is the total linear mass.)
This system determines the evolution of the temperature, pressure and
vertical velocity. Vertical fluid motions in this model are due solely
to density changes that occur in a horizontally stratified manner --- the
equations do not support acoustic or gravity waves.

The horizontal components of the momentum equation at order 1
govern the horizontal fluid flows within each material layer
corresponding to $z'=$const. Presuming that $\pnon$ yields an $O(M^2)$
correction to the leading order pressure, we may write the horizontal
components of the momentum equation as follows: Let $\mathbf{u}$
denote the horizontal components of velocity $\mathbf{v}_0$ and the subscript
$h$ denote differentiation with respect to the horizontal variables
$\mathbf{x}_h=(x^*,y^*)$. Then the horizontal momentum equation for 
$\mathbf{u}(\mathbf{x}_h,z',t')$ is 
\begin{equation}
\rho_0 \left(\dd{\mathbf{u}}{t'}
+(\mathbf{u}\cdot\nabla_h)\mathbf{u}\right)
+ \nabla_h \pnon = \frac1{\text{Re}}\left(
\mu_0 \Delta_h\mathbf{u}+ \rho_0\dd{}{z'}
\left(\rho_0\mu_0\dd{\mathbf{u}}{z'}\right) \right).
\label{32.u}
\end{equation}
Horizontal motions are incompressible, because
$\nabla_h\cdot\mathbf{u}=0$ from (\ref{32.pz}) and (\ref{32.rhoz}).
When viscosity can be neglected, then the equations corresponding to
different fluid layers decouple, and fluid layers can exhibit
arbitrary independent two-dimensional incompressible flows.

The vertical component of the momentum equation at order 1 can be used
in determining higher-order corrections of order $M^2$ for horizontally
averaged density, pressure and temperature. For the sake of brevity we omit
further discussion.

\subsection{The anelastic approximation}

In order to admit convective flows with nontrivial vertical circulation
and/or gravity waves, we can suppose that the entropy is spatially
constant at leading order. As we have indicated, this is the same as the
assumption made by Ogura and Phillips that potential temperature
variations are small compared to a reference value.

At leading order, then, the fluid density, pressure and temperature are
in hydrostatic balance and adiabatically stratified, with a (negative)
adiabatic temperature gradient.  We must neglect heat conduction to
maintain entropy constant; this restricts the time scale, and any flows
generated will be adiabatic.

To describe flows, we use the notation of section \ref{FlowRegime} for
nondimensional and leading-order quantities, and nondimensionalize
entropy according to $s=s_as^*$ where $s_a=c_{pa}\thett_a$ (as discussed
below). We postulate that the nondimensional pressure $p^*$
and entropy $s^*$ are given to order $O(M^2)$ by
\[
p^* \sim p_0(z^*)+M^2\pnon(\mathbf{x}^*,t^*), \quad
s^* \sim s_0+M^2s_1(\mathbf{x}^*,t^*),
\]
where $s_0$ is a constant.
(This neglects any fast-time acoustic corrections that may be present as
discussed in \cite{bib8}.) Then we expect the density
$\rho^* \sim \rho_0(z^*)+M^2\rho_1(\mathbf{x}^*,t^*)$, where from the
equation of state,
\begin{equation}
\rho_1 = 
\left(\frac{\partial\rho^*}{\partial p^*}\right)_{s^*} \pnon 
+\left(\frac{\partial\rho^*}{\partial s^*}\right)_{p^*} s_1 =
\frac{1}{{c_{s0}}^2}p_1 + 
(1+\thett_a\thett_0)\frac{\rho_0\alpha_{p0}}{c_{p0}}s_1.
\label{33.rho1}
\end{equation}
Here the nondimensional coefficients are evaluated at $(p_0,s_0)$.

The momentum equation at order 1 is
\begin{equation}
\rho _0\frac{D\mathbf{v}_0}{Dt^{*}} + \nabla \pnon =
-\rho_1\mathbf{\bar{g}} 
+\frac1{\text{Re}}\left(
\div(\mu _0(\nabla \mathbf{v}_0+\nabla \mathbf{v}_0^{T\,}))
+\nabla (\lambda _0\div \mathbf{v}_0)
\right).
\label{33.v0}
\end{equation}
Since heat conduction is neglected, the entropy $s_1$ is convected with
the flow, satisfying
\begin{equation}
\frac{Ds_1}{Dt^*}=0.
\label{33.s1}
\end{equation}
Because $\partial\rho_0/\partial t^*=0$, the leading-order continuity
equation yields
\begin{equation}
\div(\rho_0\mathbf{v}_0)=0.
\label{33.rho0}
\end{equation}
This constraint on the velocity should be used in solving
(\ref{33.v0}) to determine $\pnon$. Note, however, that $\pnon$ will
not be completely determined by the constraint --- Given any solution
of (\ref{33.v0}), $\pnon$ can be replaced by adding an arbitrary
time-dependent multiple of a solution
to the linearized hydrostatic balance equation
\[
\dd{\tilde p}{z^*}= -\frac{\bar{g}}{c_{s0}^2} {\tilde p}.
\]
The equations (\ref{33.v0})--(\ref{33.rho0}) 
correspond to the anelastic equations of Ogura and
Phillips, generalized for an arbitrary equation of state.

As a model for slow flows of fluids near the critical point,
these equations have some drawbacks:
First, heat conduction is entirely neglected, so the fast adiabatic
mechanism for rapid thermal response is not accounted for within this
model. Also the effect of fluid flow on thermal relaxation cannot be
evaluated.
Moreover, describing states in thermodynamic equilibrium is problematic.
At rest, the equations permit an arbitrary horizontally uniform
entropy correction $s_1$.
It may or may not be consistent with the derivation of the
equations to take this to correspond to the equilibrium entropy profile
(meaning the hydrostatic profile at constant temperature).

Regarding this point, we can ask, what is the size of the nondimensional
equilibrium entropy gradient in the regime of interest? 
To estimate this, we need to identify a typical entropy change in 
a process of interest. 
Consider a fluid at equilibrium at one temperature, subject to
a temperature change at the boundary of order $\thett_c\thett_a$.
During the early development of the thermal boundary layer, we
may suppose roughly that the entropy change in the boundary layer 
occurs at constant pressure, so it is of order $s_a=c_{pa}\thett_a$ since
$(\partial s/\partial\thett)_p=c_p/\thett$ and $\thett_a$ is small.
Then it seems reasonable to nondimensionalize the entropy by letting
$s=s_as^*$. 

Now, the equilibrium entropy gradient satisfies
\[
\frac{ds}{dz}= \left(\frac{\partial s}{\partial p}\right)_\thett
\frac{dp}{dz}= 
\frac{c_p-c_v}{\thett}
\left(\frac{\partial \thett}{\partial p}\right)_\rho (-\rho g).
\]
Nondimensionalizing this expression, we find that up to a quantity
of order one, 
\begin{equation}
\frac{ds^*}{dz^*} \sim - M^2 |\mathbf{g}^*| =
-\frac{x_a g\rho_c}{p_cp_a}.
\label{33.ds*}
\end{equation}
Under the specific conditions considered in section
\ref{FlowRegime}, $ds^*/dz^*\sim -.003$. This is moderately small but 
not quite of order $M^2$; it would make $ds_1/dz^*$ of the order of $-10^4$.  
Also, since $p_a\approx 6\thett_a$, 
the last member of (\ref{33.ds*}) shows that $ds^*/dz^*$ diverges
as temperature approaches critical, like $\thett_a^{-1}$ as $\thett_a\to0$. 

It seems that in our flow regime, the entropy gradient may not be small
enough for the anelastic approximation to be reasonable for states near
true equilibrium (at least at the level of critical density).  But
states near constant entropy do possess experimental interest.
Physically, the condition that an inviscid stratified fluid is stable
against convection is $ds/dz<0$ \cite{bib9}. In order to obtain
`pseudo-equilibrium' states with near-uniform density profiles,
ground-based experiments have been suggested on near-critical fluids
in which a steady-state heat flux is maintained so as to achieve a
marginally stable entropy profile \cite{bib12}, also see \cite{bib2}.
Note that the adiabatic density
gradient satisfies $\rho^{-1}d\rho/dz=-g/c_s^2$, which is about 1/500
m${}^{-1}$ in our flow regime, so the density is close to constant in
centimeter-sized cells. Provided that heat conduction can be neglected
on the time scale of interest, the anelastic approximation could be
useful to describe flows near such `pseudo-equilibrium' states.

\subsection{Slow flows including equilibrium}

In trying to enforce hydrostatic balance at leading order, we have found
that either flows remain strictly stratified, or equilibrium states are
not admissible as entropy must be constant at leading order in $M^2$.
In this section we return to the scaling as it was done in section
\ref{Reduce}, in which the leading-order pressure turns out to be
spatially constant.
Working in the spirit of the Boussinesq approximation, we can decide to
selectively retain some terms of higher order in $M^2$ where it would be
most useful, without affecting the formal accuracy of the system.

We propose no alteration in the equations (\ref{14t})-(\ref{14p0})
for leading-order temperature and mean pressure. (But now $p_0$
and $\thett_0$ will depend on $M^2$, through the coupling to the momentum
equation.) To order $M^2$,
supposing that the pressure is given by $p^*\sim p_0+M^2\pnon$, the
approximation to the density $\rho\approx\rho_0=\hat\rho^*(p_0,\thett_0)$
can be improved to
\[
\rho\approx\newrho \equiv \hat\rho^*(p_0+M^2\pnon,\thett_0).
\]

We propose to make this improvement only in 
equation (\ref{14v}) for the velocity. The new system of equations
governing leading-order temperature, pressure and velocity 
(neglecting capillarity) is 
\begin{eqnarray}
\frac{D\thett _0}{Dt^{*}}&=&
\left( 1-\Gamma ^{-1}\frac{c_{v0}}{c_{p0}}\right)
\frac{K_{T0}}{\alpha _{p0}}\frac{dp_0}{dt^{*}} +
\frac{D_T}{\rho_0c_{p0}}\div(\kappa _0\nabla \thett _0),  \tgg{34t}\\
K_{T0}\frac{dp_0}{dt^{*}}&=&\alpha _{p0}\frac{D\thett _0}{Dt^{*}}-
\div\mathbf{v}_0  \tgg{34p0} , \\
\newrho\frac{D\mathbf{v}_0}{Dt^*} +\nabla \pnon &=&
-\newrho\mathbf{g^*}+ \frac1{\text{Re}}
\left( \div(\mu_0(\nabla\mathbf{v}_0 + \nabla\mathbf{v}_0^T))
+\nabla(\lambda_0\div\mathbf{v}_0) \right).
\label{34.v0}
\end{eqnarray}
The temperature evolution and continuity equations
(\ref{34t})--(\ref{34p0}) may be replaced by the equivalent pair
(\ref{32.s})--(\ref{32.p}), in which the material derivative
$Dp_0/Dt^*$ can be replaced by the ordinary derivative $dp_0/dt^*$.

As with the anelastic equations, the pressure correction $\pnon$
should be determined in solving (\ref{34.v0}) to satisfy the implied
constraint on the divergence of velocity from (\ref{34p0}).
The pressure correction will not be unique, but any two solutions
$\pnon$ and $\tilde\pnon$ that correspond to the same
$(p_0,\thett_0,\mathbf{v}_0)$ will be related by 
\begin{equation}
\nabla p_1+\hat\rho^*(p_0+M^2p_1,\thett_0)\mathbf{g}^* = 
\nabla \tilde p_1+\hat\rho^*(p_0+M^2\tilde p_1,\thett_0)\mathbf{g}^* .
\end{equation}
Therefore the difference $p_1-\tilde p_1$ is a function of $z^*$ and
$t^*$ that is determined solely in
terms of a function of $t^*$ by solving an ordinary differential equation. 

The system consisting of equations (\ref{34t})--(\ref{34.v0})
can be reformulated to better reveal its evolutionary
character exactly as in section \ref{Reform}. In dimensional form with
pressure $\pzero(t)+\pdim(x,t)$, 
where $\pzero=p_c(1+p_ap_0)$ and $\pdim=p_cp_aM^2\pnon$, one
obtains exactly the system (\ref{15p})--(\ref{15divw}) except that
(\ref{15w}) is replaced by
\begin{eqnarray}
\newrho \frac{D\mathbf{w}}{Dt} &=&-\nabla \pdim
 -\newrho\mathbf{g}-\newrho \frac{D(\nabla \phi )}{Dt}
\nonumber\\
 &&+\div\left(\mu (\nabla (\mathbf{w}+\nabla \phi )+
 \nabla (\mathbf{w}+ \nabla \phi )^{T})\right)+
 \nabla (\lambda \Delta \phi ),  \tgg{34w}
\end{eqnarray}
where the (now dimensional) density $\newrho$ is determined from
pressure and temperature by the equation of state:
$\newrho=\hat\rho(\pzero+\pdim,\thett)$. 
We anticipate that, like system (\rf{15p})--(\rf{15divw}), solutions
of (\ref{34t})--(\ref{34.v0}) are determined by initial values for the
temperature field, mean pressure, and divergence-free part of the
velocity field.

The system (\ref{34t})--(\ref{34.v0})
admits as rest states true equilibrium states with constant temperature
and hydrostatic balance between the total pressure $p_0+M^2\pnon$ and the
improved density $\newrho=\hat\rho^*(p_0+M^2\pnon,\thett_0)$.
It filters acoustic waves but admits gravity waves, as we shall show
in the subsection to follow.
Moreover, heat conduction need not be neglected, so the adiabatic effect
can be modeled.
It should be interesting to study what flows are
generated when large density changes in the thermal boundary layer
are present together with very stable equilibrium entropy profiles away
from the boundary.

We remark that it is evidently not necessary to replace $\rho_0$ by
$\newrho$ in the acceleration term of (\ref{14v}) to gain 
true equilibria as rest states.  In some circumstances it may be more
convenient not to make this replacement. But it turns out
that equation (\ref{34.v0}) is slightly more convenient when we study
the linearized equations for gravity waves; see the next section.

The system (\ref{34t})--(\ref{34.v0}) certainly has
shortcomings. First, in the flow regime described in section
\ref{FlowRegime}, $|\mathbf{g}^*|=gt_a^2/x_a$ is still rather large.
This problem diminishes if a larger space scale or smaller time scale
is relevant.
The system can be expected to lose formal validity if solutions become
large or singular, as may well happen in a nonadiabatic convection
process. 
Another point is that while the leading-order total energy is
conserved in time for the weak-gravity equations
(\ref{15p})--(\ref{15divw}), this is not strictly true for the system
(\ref{34t})--(\ref{34.v0}).  The time derivative of the
leading-order total energy is formally of order $M^2$ instead.

\subsection{Gravity waves in the linear approximation}

We wish to verify that in the linear approximation near a stably
stratified rest state for which entropy decreases with height, the
system (\ref{34t})--(\ref{34.v0}) admits gravity waves but not
acoustic waves when heat conduction and viscosity are neglected. We
shall also show that, in the special case of a perfect gas with an
exponential density profile and constant sound speed, in the limit of
large wave number the gravity-wave frequency approaches the
Brunt-\Vaisala frequency $N$ corresponding to a compressible
fluid. This frequency satisfies
\begin{equation}
N^2 = -\frac{g}{\rho_e}\frac{d\rho_e}{d z} -
\frac{g^2}{c_s^2} .
\label{35bv}
\end{equation}
The second term does not appear in the usual treatment of gravity
waves for a stratified incompressible fluid, in which the density is
advected with the flow, cf.\ \cite{Yih}.

When heat conduction and viscosity are neglected in the system 
(\ref{34t})--(\ref{34.v0}), the leading-order pressure
is constant in time as well as space. Consequently the temperature
(and entropy) are advected with the flow, and the velocity field has
zero divergence. In dimensional form with $p=\pzero+\pdim$ and
$\newrho=\tilde\rho(p,s)$, the governing equations take the form
\begin{eqnarray}
\frac{Ds}{Dt}&=&0, \label{35s}\\
\div \mathbf{v}& =& 0, \label{35div}\\
\newrho \frac{D\mathbf{v}}{Dt} +\nabla p &=&-\newrho \mathbf{g}.\label{35v}
\end{eqnarray}
Near a rest state where $(p,s,\mathbf{v})=(p_e(z),s_e(z),\mathbf{0})$,
we write 
\[
p=p_e+\tilde{p}, \quad s=s_e+\tilde{s},\quad
\mathbf{v}= (\tilde{u},\tilde{v},\tilde{w})
\]
and $\rho_e(z)=\tilde\rho(p_e,s_e)$. Then we linearize, obtaining
\begin{eqnarray*}
\frac{\partial \tilde{s}}{\partial t}+\tilde{w} s_e'(z) &=&0, \\
\div\mathbf{v}&=&0,\\
\rho_e \frac{\partial\mathbf{v}}{\partial t} +\nabla\tilde{p}
&=& - \left(
 \left(\frac{\partial\rho}{\partial p}\right)_s\tilde{p} + 
\left(\frac{\partial\rho}{\partial s}\right)_p
\tilde{s}\right)\mathbf{g} ,
\end{eqnarray*}
where the coefficients
are evaluated at $(p_e,s_e)$. 
Let us suppose that periodic boundary conditions are specified in the
horizontal variables $(x,y)$. We look for normal modes with
\[
(\tilde{p},\tilde{s},\tilde{u},\tilde{v},\tilde{w}) = 
e^{i(k_1x+k_2y-\omega t)}(\bar{p},\bar{s},\bar{u},\bar{v},\bar{w})(z),
\]
and eliminate $\bar{p}$, $\bar{s}$ and the horizontal components
of velocity from the system. Note that the sound speed satisfies
$c_s^{-2}=(\partial\rho/\partial p)_s$, and define
\begin{equation}
\alpha(z)= -\frac1{\rho_e} \frac{d\rho_e}{dz}, \quad
\beta(z)= -\frac1{\rho_e}\left(\frac{\partial \rho}{\partial s}\right)_p 
\frac{ds_e}{dz} = \alpha(z)-\frac{g}{c_s^2}.
\end{equation}
(The last identity holds due to hydrostatic balance.)
Then for the vertical velocity component
$\wbar(z)$ we obtain the equation
\begin{equation}
-\frac{d^2\wbar}{dz^2}+\beta \frac{d\wbar}{dz}+ \left(
k^2-\frac{k^2\beta g}{\omega^2} \right)\wbar = 0,
\label{35wbar}
\end{equation}
where $k^2=k_1^2+k_2^2$. The vertical velocity must vanish at the top and
bottom boundaries. So, given a horizontal wave number $k$, possible 
oscillation frequencies $\omega$ are determined by solving the
eigenvalue problem in (\ref{35wbar}) with Dirichlet boundary
conditions.  

We note that in the special case of perfect gas at constant
temperature, the density profile is
exponential and sound speed is constant, and so $\alpha(z)$ and 
$\beta(z)$ are constant. Then 
(\ref{35wbar}) has explicit solutions of the form
$\wbar(z)=e^{\beta z/2}\sin(nz)$, whence the gravity-wave dispersion
relation is given by 
\begin{equation}
\omega^2 = \frac{k^2 \beta g}{k^2+n^2+\frac14 \beta^2}.
\label{35disp1}
\end{equation}
In the limit $k^2\to\infty$, this expression approaches $\beta g=N^2$,
where $N$ from (\ref{35bv}) is the Brunt-\Vaisala frequency for a
compressible fluid.  For comparison, for a fully compressible fluid in
which one starts with $D\newrho/Dt+\newrho\div\mathbf{v}=0$ in place
of (\ref{35div}), the dispersion relation in this special case is
\[
\omega^2 = \frac{c_s^2}2\left(
k^2+n^2+\frac{\alpha^2}4 \pm \sqrt{
\left(k^2+n^2+\frac{\alpha^2}4 \right)^2 -4k^2\frac{N^2}{c_s^2} }\right).
\]
For large $k^2$, the plus sign yields $\omega^2\approx c_s^2k^2$,
corresponding to acoustic waves, and the minus sign yields
$\omega^2\approx N^2$, corresponding to gravity waves.
The magnitude of any oscillation frequency
$\omega$ that satisfies (\ref{35disp1}) is less than $N$, showing that
the system (\ref{35s})--(\ref{35v}) supports gravity waves but not
acoustic waves.

In general, when the coefficients in (\ref{35wbar}) vary with $z$, we
can obtain an upper bound on oscillation frequencies as follows.  Let
$q(z)=\exp\left(-\int^z\beta(\zeta)\,d\zeta\right)$, multiply
equation (\ref{35wbar}) by $q\wbar$ and integrate over $z\in[0,L]$,
from bottom to top. One obtains
\[
\int_0^L q(z)\wbar'(z)^2\,dz + \frac{k^2}{\omega^2}
\int_0^L (\omega^2-\beta g)q(z)\wbar(z)^2\,dz = 0.
\]
Since the second integrand cannot be everywhere
nonnegative, it follows that 
\begin{equation}
\omega^2 < \max_z \beta(z) g .
\label{35omega}
\end{equation}

The dispersion relation in (\ref{35disp1})  is very similar to that
obtained in the usual case of an incompressible fluid when one assumes
the density is advected with the flow. Starting from the governing
equations
\begin{eqnarray}
\frac{D\rho}{Dt}&=&0,\\
\div \mathbf{v}& =& 0, \\
\rho \frac{D\mathbf{v}}{Dt} +\nabla p &=& - \rho \mathbf{g},
\end{eqnarray}
one finds in similar fashion that the equation corresponding to 
(\ref{35wbar}) is
\begin{equation}
-\frac{d^2\wbar}{dz^2}+\alpha \frac{d\wbar}{dz}+ \left(
k^2-\frac{k^2\alpha g}{\omega^2} \right)\wbar = 0,
\label{35wbar2}
\end{equation}
and for an exponential density profile the dispersion relation is 
\begin{equation}
\omega^2 = \frac{k^2 \alpha g}{k^2+n^2+\frac14 \alpha^2}.
\label{35disp2}
\end{equation}
In the limit $k^2\to\infty$ this approaches $\alpha g=N_0^2$, where $N_0$ is
the {\it usual} Brunt-\Vaisala frequency for an incompressible fluid.
In many circumstances the difference between $N_0$ and $N$ may be
negligible, but it is interesting that the dispersion relation arising
from the system (\ref{35s})--(\ref{35v}) more faithfully approximates
the compressible case in this respect.

\subsection{Final remarks}

We close with a few remarks intended to clarify the differences
between the new system (\ref{34t})--(\ref{34.v0}) and the anelastic
system (\ref{33.rho1})--(\ref{33.rho0}). For purposes of comparison,
we neglect heat conduction. In this case, the new system
(\ref{32.s}), (\ref{32.p}), (\ref{34.v0}) becomes
\begin{eqnarray}
\frac{Ds_0}{Dt^*} &=& 0, \label{36s}\\
\div\mathbf{v}_0 &=& 0, \label{36div}\\
\newrho\frac{D\mathbf{v}_0}{Dt^*} +\nabla \pnon &=&
-\newrho\mathbf{g^*}+ \frac1{\text{Re}}
\left( \div(\mu_0(\nabla\mathbf{v}_0 + \nabla\mathbf{v}_0^T))
+\nabla(\lambda_0\div\mathbf{v}_0) \right), \label{36v0}
\end{eqnarray}
where $\newrho=\tilde\rho^*(p_0+M^2p_1,s_0)$.

For each system, the source of the constraint on velocity ((\ref{36div})
or (\ref{33.rho0}) respectively) is the continuity equation 
\begin{equation}
\frac{D\rho_0}{Dt^*}+\rho_0 \div\mathbf{v}_0 = 0,
\label{36cont0}
\end{equation}
where $\rho_0=\tilde\rho^*(p_0,s_0)$. For the anelastic system, the
leading order entropy $s_0$ is constant and the pressure $p_0$ is a
function of $z$ determined by the hydrostatic balance equation
(\ref{32.hydro}), so $D\rho_0/Dt^*=w_0\partial\rho_0/\partial
z^*$. Using hydrostatic balance we can also write (\ref{33.rho0}) in
the form essentially given by Batchelor \cite{B}:
\begin{equation}
\div\mathbf{v}_0 - w_0 \frac{\bar{g}}{c_{s0}^2} = 0.
\label{36cont1}
\end{equation}

For the new system (\ref{36s})--(\ref{36v0}), since
$p_0$ is a constant when heat conduction is neglected,
we have $D\rho_0/Dt^*=0$ and this is
why the velocity field is divergence-free. Note that if we seek to
`improve' equation (\ref{36cont0}) by replacing $\rho_0$ by
$\newrho=\tilde\rho^*(p_0+M^2p_1,s_0)$, then we recover the original fully
compressible system without simplification! Indeed, the essential difference
between the fully compressible system (\ref{13.1v})--(\ref{13.1p}) 
(neglecting heat conduction and viscous power terms), and
system (\ref{36s})--(\ref{36v0}) with $p^*=p_0+M^2p_1$,
is precisely that in the new system a term proportional to
$M^2Dp_1/Dt^*$ is neglected in the continuity equation.

This point suggests a modification to the system
(\ref{36s})--(\ref{36v0}) in a situation with possible relevance for
atmospheric circulations. Suppose gravity is rather strong but the pressure
correction $p_1$ does not happen to deviate significantly (more than
$O(1)$) from some time-independent reference state $\bar{p}(z^*)$
that determines a reference density profile $\bar\rho(z^*)$ via a 
hydrostatic balance equation
\begin{equation}
\nabla \bar{p} = -\bar\rho(z^*) \mathbf{g}^*.
\label{36hyd}
\end{equation}
Then we replace $\rho_0$ in (\ref{36cont0}) by 
$\bar\rho_0=\tilde\rho^*(p_0+M^2\bar{p},s_0).$ Note that
$\bar\rho_0$ can depend on $(x^*,y^*,t^*)$ as well as $z^*$ through
$s_0$. Since $M^2D\bar{p}/Dt^*=M^2w_0(d\bar{p}/dz^*)$, 
the continuity equation becomes
\begin{equation}
\div\mathbf{v}_0 - w_0 \frac{\bar{g}}{\bar{c}_{s0}^2} 
\frac{\bar\rho}{\bar\rho_0} = 0,
\label{36cont2}
\end{equation}
where the coefficient $\bar{c}_{s0}^2$ is evaluated at
$(p_0+M^2\bar{p},s_0)$ and we have used $M^2 g^*=\bar{g}$. 
This equation replaces (\ref{36div}), without changing the formal
validity of the approximation.

\bigskip
\textbf{Acknowledgments:}
This work was partially supported by the National Science Foundation
under grants DMS 94-03871 and DMS 97-04924. The second author also
acknowledges the support of Krispin Technologies under NASA SBIR contract
NAS3-97087. We thank H. Boukari and R. Gammon for discussions related
to this work, and thank R. McLaughlin for recommending reference \cite{OP}.


\pagebreak




\begin{thebibliography}{99}

\bibitem{AMW}  D. M. Anderson, G. B. McFadden and A. A. Wheeler,
\textit{Diffuse-interface methods in fluid mechanics}, Ann. Rev. Fluid 
Mech. 30, 139--165 (1998).

\bibitem{B} G. K. Batchelor, \textit{The conditions for dynamical
similarity of motions of a frictionless perfect-gas atmosphere}, 
Quart. J. Roy. Meteo. Soc. 79, 224--235 (1953).

\bibitem{bib1}  R. F. Berg, \textit{Thermal equilibration near the critical
point: Effects due to three dimensions and gravity}, Phys. Rev. E 48, 
1799--1805 (1993).

\bibitem{bib2}  H. Boukari, M. E. Briggs, J. N. Shaumeyer, and R. W. Gammon,
\textit{Critical speeding up observed}, Phys. Rev. Lett.65,
2654--2657 (1990).

\bibitem{bib3}  H. Boukari, R. Pego, and R. W. Gammon, \textit{Calculation of 
the dynamics of gravity-induced density profiles near a liquid-vapor critical
point}, Phys. Rev. E 52, 1614--1625 (1995).

\bibitem{bib4}  H. Boukari, J. N. Shaumeyer, M. E. Briggs and R. W. Gammon,
\textit{Critical speeding up in pure fluids}, Phys. Rev. A 41,
2260--2263 (1990).

\bibitem{bib5}  A. J. Chorin and J. E. Marsden, \textit{A Mathematical
Introduction to Fluid Mechanics}, 3rd. ed., Springer-Verlag, New
York, 1993.

\bibitem{DP2} D. L. Denny and R. L. Pego, 
\textit{Solutions for a model of low-speed flow for highly compressible 
fluids with capillary effects}, in preparation.

\bibitem{bib6}  J. E. Dunn and J. Serrin, \textit{On the thermomechanics of
interstitial working}, Arch. Rational Mech. Anal. 88, 95--133 (1985).

\bibitem{D} D. Durran, \textit{Improving the anelastic approximation},
J. Atmos. Sci. 46, 1453--61 (1989).

\bibitem{DF} J. A. Dutton and G. H. Fitchl, \textit{Approximate
equations of motion for gases and liquids}, J. Atmos. Sci. 26, 241--54
(1969). 

\bibitem{bib7}  P. Embid, \textit{Well-posedness of the Nonlinear Equations 
for Zero Mach Number Combustion}, Ph. D. Thesis, University of California,
Berkeley, 1984.

\bibitem{Zeno} 
R. Gammon, personal communication.\quad Also see the ZENO home page at the 
URL {\tt http://roissy.umd.edu/}\ . The experimental design is
described at the URL
{\tt http://roissy.umd.edu/usmp3/reminder.html}\ . 

\bibitem{IKW1} C. Ikier, H. Klein and D. Woermann, \textit{Optical
observation of the gas/liquid phase transition in near-critical
SF${}_6$ under reduced gravity}, J. Coll. Int. Sci. 178, 368--70 (1996).

\bibitem{IKW2} C. Ikier, H. Klein and D. Woermann, \textit{Density
equilibration in a near-critical fluid under reduced gravity},
Ber. Bunsenges. Phys. Chem. 100 (8), 1308-11 (1996).

\bibitem{bib8}  S. Klainerman and A. Majda, \textit{Singular Limits of 
Quasilinear Hyperbolic Systems with Large Parameters and the Incompressible 
Limit of Compressible Fluids}, Comm. Pure Appl. Math. 34, 481--524 (1981).

\bibitem{KZM} A. B. Kogan, F. Zhong and H. Meyer, 
\textit{Dynamics of density equilibration near the liquid-vapor 
critical point of He-3}, Czechoslovak J. Phys. 46, Suppl. 1, 71-2 (1996).

\bibitem{bib9}  L. D. Landau and E. M. Lifshitz, \textit{Fluid Mechanics}, 
2nd. ed., Pergamon, Oxford, 1987.

\bibitem{LH} F. B. Lipps and R. S. Hemler, 
\textit{A scale analysis of deep moist
convection and some related numerical calculations}, J. Atmos. Sci.
39, 2192--2210 (1982).

\bibitem{bib10}  A. Majda, \textit{Compressible Fluid Flow and Systems of
Conservation Laws in Several Space Variables}, Applied Mathematical Sciences
Vol. 53, Springer, New York, 1984.

\bibitem{bib11}  A. Majda and J. Sethian, \textit{The derivation and numerical
solution of the equations for zero Mach number combustion}, Combust. Sci.
Tech. 42, 185--205 (1985).

\bibitem{bib12}  M. R. Moldover, J. V. Sengers, R. W. Gammon and J. R. Hocken,
\textit{Gravity effects in fluids near the gas-liquid critical point}, 
Rev. Mod. Phys. 51, 79--99 (1979).


\bibitem{OP} Y. Ogura and N. A. Phillips, \textit{Scale analysis of
deep and shallow convection in the atmosphere}, J. Atmos. Sci. 19,
173--9 (1962). 

\bibitem{bib14}  A. Onuki and R. A. Ferrell, \textit{Adiabatic heating effect 
near the gas-liquid critical point}, Physica A 164, 245--264 (1990).

\bibitem{bib15}  A. Onuki, H. Hao, and R. A. Ferrell, \textit{Fast adiabatic
equilibration in a single-component fluid near the liquid-vapor critical
point}, Phys. Review A 41, 2256--2259 (1990).

\bibitem{bib16}  V. A. Rabinovich, \textit{Thermophysical Properties of
Neon, Argon, Krypton, and Xenon}, Hemisphere Publishing Corp., New York,
1988.

\bibitem{bib17} R. G. Rehm and H. R. Baum, \textit{The equations of
motion for thermally driven, buoyant flows},
J. Res. Natl. Bur. Stand. 83, 297--308 (1973).

\bibitem{bib18}  J. S. Rowlinson and  B. Widom, \textit{Molecular Theory of
Capillarity}, Clarendon Press, Oxford, 1982.

\bibitem{B28} J. V.  Sengers, \textit{Transport properties of fluids
near critical points}, Int. J. Thermophys. 6, 203--232 (1985). 

\bibitem{bib19}  J. V. Sengers, R. S. Basu and J. M. H. Levelt Sengers,
\textit{Representative equations for the thermodynamic and transport 
properties of fluids near the gas-liquid critical point}, NASA Contractor 
Report 3424, 1981.

\bibitem{B27}  H. L. Swinney and D. L. Henry, \textit{Dynamics of fluids near 
the critical point: decay rate of order-parameter fluctuations},
Phys. Rev. A 8, 2586--2617 (1973), and references therein.


\bibitem{WO} R. Wilhelmson and Y. Ogura, \textit{The pressure
perturbation and the numerical modeling of a cloud},
J. Atmos. Sci. 29, 1295--1307 (1972).

\bibitem{Yih} C.-S. Yih, \textit{Stratified Flows}, Academic Press,
New York, 1980.

\bibitem{ZC} B. Zappoli and P. Carles,
\textit{The thermo-acoustic nature of the critical speeding up},
 Eur. J. Mech. B/Fluids 14, 41--65 (1995).

\bibitem{Zetal} B. Zappoli, S. Amiroudine, P. Carles and J. Ouazzani,
\textit{Thermoacoustic and buoyancy-driven transport in a square side heated
cavity filled with a near critical fluid}, J. Fluid Mech. 316, 53-72 (1996).

\bibitem{bib21}  F. Zhong and H. Meyer, \textit{Density equilibration near 
the liquid-vapor critical point of a pure fluid: Single phase}, 
Phys. Rev. E 51 (1995) 3223--3241.

\end{thebibliography}
\end{document}